\documentclass[11pt, oneside]{article}

\usepackage{setspace}
\usepackage{lmodern}
\usepackage[english]{babel}
\usepackage[round]{natbib}
\usepackage{etoolbox}
\usepackage[utf8]{inputenc}
\usepackage{endnotes}
\usepackage{amsmath, empheq}
\usepackage{amssymb}
\usepackage{amsthm}
\usepackage{amscd}
\usepackage{mathrsfs}
\usepackage{mathtools}
\usepackage{dsfont}
\usepackage{thmtools}
\usepackage[labelfont={small,bf}, font={footnotesize,sf,doublespacing},
			justification=RaggedRight, singlelinecheck=false]{caption}
\usepackage{array}
\usepackage{booktabs}
\usepackage{geometry}
\usepackage{chngcntr}

\newcommand{\km}{\texttt{km} }
\newcommand{\hab}{\texttt{hab} }
\newcommand{\R}{\mathds{R}}
\newcommand{\dif}[1]{\,\mathtt{d}#1}
\newcommand{\E}[1]{\,10^{#1}}
\newcommand{\guil}[1]{``#1''}
\newcommand{\captionLine}{
	\vspace{0\baselineskip}
	\hrule
	\smallskip}
\apptocmd{\thebibliography}{\raggedright}{}{}
\author{}

\title{\textbf{}}

\date{\textsuperscript{1}{Université catholique de Louvain, Center for Operations Research and Econometrics, Belgium}\\
\textsuperscript{2}{Université du Luxembourg, Institute of Geography and Spatial Planning, Luxembourg}\\
\textsuperscript{3}{Luxembourg Institute of Socio-Economic Research, Luxembourg}\\
\bigskip
\textsuperscript{*}{Corresponding author: \textsf{justin.delloye@uclouvain.be}}}

\begin{document}

\thispagestyle{empty}
\makeatletter
	{\huge\bfseries\noindent
		Alonso and the Scaling of Urban Profiles
	\par}
	\vspace{0.5\baselineskip}
	{\LARGE\noindent
		Justin Delloye\textsuperscript{1}, Rémi Lemoy\textsuperscript{2} and Geoffrey Caruso\textsuperscript{2,3}
	\par}
	\vspace{1\baselineskip}
	{\large\noindent
		\textsuperscript{1}{Université catholique de Louvain, Center for Operations Research and Econometrics, Belgium}\\
\textsuperscript{2}{Université du Luxembourg, Institute of Geography and Spatial Planning, Luxembourg}\\
\textsuperscript{3}{Luxembourg Institute of Socio-Economic Research, Luxembourg}
	\par}
	\vspace{1\baselineskip}
	{\noindent
		Correspondence: Justin Delloye, Center for Operations Research and Econometrics, Voie du Roman Pays, 34 - L1.03.01, 1348 Louvain-la-Neuve, Belgium\\E-mail: \texttt{justin.delloye@uclouvain.be}

	\par}
\makeatother
\vspace{5mm}

\vspace{\baselineskip}

\renewcommand{\abstractname}{\vspace{-\baselineskip}}
\begin{abstract}
\noindent \textit{The scaling of urban characteristics with total population has become an important research field since one needs to better understand the challenges of urban densification. Yet urban scaling research is largely disconnected from intra-urban structure. In contrast, the monocentric model of Alonso provides a residential choice-based theory to urban density profiles. However, it is silent about how these profiles scale with population, thus preventing empirical scaling studies to anchor in a strong micro-economic theory. This paper bridges this gap by introducing power laws for land, income and transport cost in the Alonso model. From this augmented model, we derive the conditions at which the equilibrium urban structure matches recent empirical findings about the scaling of urban land and population density profiles in European cities. We find that the Alonso model is theoretically compatible with the observed scaling of population density profiles and satisfactorily represents European cities. This compatibility however challenges current empirical understanding of wage and transport cost elasticities with population, and requires a scaling of the housing land profile that is different from the observed. Our results call for revisiting theories about land development and housing processes as well as the empirics of agglomeration benefits and transport costs.}
\end{abstract}

%%%%%%%%%%%%%%%%%%%%%%%%%%%%%%%%%%%%%%%%%%%%%%%%%%%%%%%%%%%%%%%%%%%%%%%%%%%%%%
\section*{Introduction}
It is theoretically elegant and empirically convenient to think of all the good and bad of cities simply in terms of their total population. We live in an increasingly urban World \citep{UNHABITAT2016} and liaising the social and environmental outcomes of cities to their size is definitely an important question today and for tomorrow. Yet, we know that many outcomes of cities depend crucially on their internal structure, especially on how densely citizens occupy the land they have developed. This occupation emerges from the location decisions of many people interacting in space and is often described or discussed in radial terms, that is how far reaching a city is (the urban fringe distance) and how flat/steep its density profile is. This is a key interest of theoretical and empirical urban economics \citep[see][for a reminder]{anas_urban_1998} and the favourite playground of urban planning. The long dispute between compactness or sprawl \citep[e.g.][for a quick summary]{ewing2014compactness} just shows how much this internal structure matters and is worth being studied. Therefore, before summing-up a city as the outcome of a single termed function of population, one needs first to make sure that the internal structure of cities is independent of population, or is at least independent of a simple (well-behaved) transformation of population, and second -- particularly if desirable actions need to be made with potential social impacts -- one needs to know if this internal structure responds to the same underlying decisional processes independent of size, in other terms that the same urban theory holds across the size distribution of cities.

\citet{nordbeck1971urban} provided an intuition to the first need, and opened up a literature strand on allometric urban growth by assuming that cities, similarly to biological objects, keep the same form across size. \citet{lemoy_scaling_2017} recently endorsed this idea and empirically identified the homothetic transformations of density and land profiles with population for European cities. A logical extra step is then to address the second need described above and assess whether models that can generate observed urban radial profiles can also replicate their scaling with population. Finding a valid model that can be applied to any cities after simple rescaling would definitely bear powerful implications for understanding cities and identifying generic planning recipes independent of size. The Alonso-Muth-Mills monocentric framework \citep{alonso_location_1964, muth_cities_1969, mills_studies_1972} is a perfect candidate because it issues micro-foundations to urban expansion limits and density gradients. It does so after fixing population in its closed equilibrium form, or after fixing its social outcome (utility) in its open form where equilibrium with other cities is then assumed and the population an output.

In this paper we assess the theoretical ability and conditions for the Alonso model to replicate the scaling behaviour of urban density and urban land profiles. Given that the Alonso model however assumes a fully urbanised disc, which is inconsistent with the presence of semi-natural land within cities and with a decreasing profile of urbanised land, our model exogeneously relaxes this assumption. We then test how the standard form of the Alonso model and its relaxed land use form (named ``Alonso-LU'') empirically perform in Europe after a parsimonious calibration calling only three parameters.

%%%%%%%%%%%%%%%%%%%%%%%%%%%%%%%%%%%%%%%%%%%%%%%%%%%%%%%%%%%%%%%%%%%%%%%%%%%%%%
\section*{Background}
In the last few decades, and particularly since the advent of the complexity paradigm  \citep{arthur_economy_1997, vicsek_complexity:_2002, batty_cities_2007, white_modeling_2015}, researchers have reinvested the question of scaling patterns for cities. Most of these investigations, conducted by economists, physicists and geographers, have been dedicated to systems of cities, i.e. the inter-urban scale, with particular attention on rank-size distributions and empirical testing of Zipf's law through space and time \citep[e.g.][]{pumain_scaling_2004, bettencourt_growth_2007, shalizi_scaling_2011, batty_new_2013, louf_scaling:_2014, leitao_is_2016,cura2017old}. Theoretical grounds have been provided along dissipative systems analogies  \citep{bettencourt_origins_2013}  or Gibrat's law of proportionate growth \citep{pumain_dynamique_1982, gabaix_zipfs_1999}, ruling out economics of agglomeration. These studies are essentially \textit{a-spatial}, meaning that cities could be reshuffled anywhere \citep[except for instance][]{pumain_urban_2017} and, most importantly in light of our objectives, meaning that their intra-urban structure is ignored.

Geographers and physicists have also explored intra-urban scaling, especially \citet{batty_fractal_1994, frankhauser_fractalite_1994} have initiated research on fractal geometries and identified their resemblance with land urbanisation patterns. Most of this literature is devoted to identifying irregular urban boundaries \citep[e.g.][]{tannier_fractal_2011} and non-monocentric patterns \citep{chen_fractal_2013}. Apart from two noticeable exceptions by \citet{cavailhes_where_2004, cavailhes_residential_2010}, no link is explicitly drawn however in the fractal literature with the fundamental location trade-offs of the urban economic tradition. Even in these particular exceptions, though, densities and rents are output on top of an exogenous land pattern, either multi-fractal or inspired by a Sierpinski carpet. Furthermore, despite fractality implies repeating structures across scales, this literature does not relate to city size distribution and inter-urban research.

In urban economics, the set of monocentric models arising from Alonso-Muth-Mills explicitly aim at explaining land use patterns, densities and land/housing markets as a function of distance to an exogenous Central Business District (CBD) \citep{fujita_urban_1989} and a large theoretical literature has emerged \citep{fujita_economics_2013, duranton_handbook_2015}. Some links have been drawn with inter-city research and the distribution of cities but without addressing population scaling as such. It is rather focused on agglomeration effects and migration costs between cities \citep[e.g.][]{tabuchi_number_2005}. Empirical studies are less numerous \citep{cheshire_handbook_1999, ahlfeldt_if_2008, spivey_mills-muth_2008} and again hardly focus on scaling properties with respect to population size. A notable exception is \cite{mcgrath_more_2005} who, following \cite{brueckner_economics_1983-1}, studied the evolution of city size (measured as the area or radius of urban regions) with different parameters, including population, using data from 33 U.S. cities over five decades. He observed that the sign of the variation of city size is statistically consistent with urban economic models, but did not develop the exact relationship nor the scaling properties of the land or density profiles.

\paragraph{} Overall, population scaling in inter-urban research stays strongly disconnected from intra-urban empirics and theory. Scaling laws consider averaged attributes while ignoring the making of urban patterns and their effects on these attributes. They especially ignore the fundamental trade-off between transport and land/housing costs within cities as documented after Alonso, that gives rise to decreasing population and urban land density profiles with distance to the CBD. We attempt to bridge this theoretical gap by integrating recent empirical hints from \citet{lemoy_scaling_2017} about the scaling of urban profiles into the Alonso model. 

\begin{figure}
\centering
\includegraphics{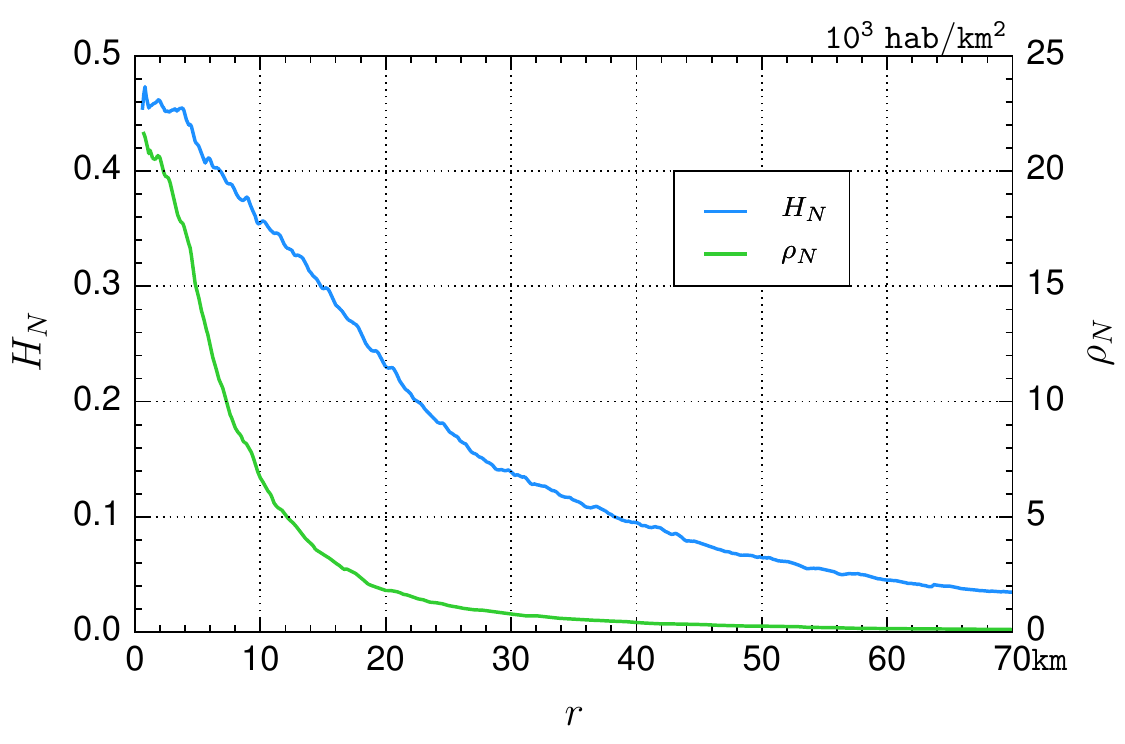}
\captionLine
\caption{\textbf{Average share of housing land and population density profiles.} These profiles have been rescaled without loss of generality to London's population (the largest European Larger Urban Zone in 2006), taken as $N=1.21\E{7}$ \citep[see][]{lemoy_scaling_2017}.}
\label{fig_empProf}
\end{figure}

\cite{lemoy_scaling_2017} carried out a radial analysis for over 300 European cities of more than $100\,000$ inhabitants as of 2006. They analysed the profile of the share of land devoted to housing with distance to the CBD and found that all profiles superpose after their abscissa is rescaled with respect to urban population, thus following a two-dimensional (horizontal) homothetic scaling. Similarly, they analysed population density profiles and found these superpose after a rescaling in abscissae and ordinates, thus following a three-dimensional homothetic scaling. Optimal rescaling is obtained numerically with the square root of population for land use profiles and the cube root of population for population density profiles. This yields the generic profiles shown on Fig. \ref{fig_empProf}, with $H_{N}(r)$ the share of housing land and $\rho_{N}(r)$ the population density as a function of distance $r$ to the CBD. These are representative profiles which can be rescaled to describe any European city once its population $N$ is given.

The validity of Fig. \ref{fig_empProf} across city sizes cannot be explained by previous geographical research in scaling laws because it has not been linked to a radial intra-urban approach so far. In order to be explained by the standard monocentric theory, one then needs to introduce scaling laws in the Alonso framework before assessing how it suits empirical evidence. By doing so we actually start bridging the gap between intra-urban and inter-urban theory.

In addition, we see from Fig. \ref{fig_empProf} that the land used for housing is far from the constant share (usually 100\%) assumed by the Alonso model. In Europe, at the CBD, land for housing is actually about half of the land and this share decreases to reach only 10\% at 40 km of the CBD for cities like London or Paris. At this stage of the research and given our primary focus on scaling, we opt for an exogenous treatment of the housing land development process. We are aware of models that permit non-urbanised land (agricultural or semi-natural) to be interspersed within the urban footprint because of spatial interactions with residents \citep{cavailhes_periurban_2004, caruso_spatial_2007}  but leave their integration to future work.

\paragraph{} We organize the remainder of the paper into a theoretical section and an empirical one. In the next section, we introduce power laws for density and for housing land profiles in a relaxed version of the Alonso model where housing does not necessarily fully occupy land around the CBD. We then derive conditions at which the equilibrium profiles match the scaling exponents of \citet{lemoy_scaling_2017}. In another, empirical section, we use their European data to calibrate the model, respectively its standard form with constant occupation of land (Alonso) and the relaxed version with exogenously given land profile function (Alonso-LU), thus leaving the model to produce densities within these constraints. We conclude in the last section.

%%%%%%%%%%%%%%%%%%%%%%%%%%%%%%%%%%%%%%%%%%%%%%%%%%%%%%%%%%%%%%%%%%%%%%%%%%%%%%
\section*{Theory}

First, we define the setting and introduce homothetic scaling in density and housing land profiles. Second, we define the decision making of households and introduce scaling for parameters of this choice (income and transport cost). Third, we take an intra-urban perspective, and resolve the equilibrium for the closed form (given population, endogenous utility) of the Alonso model with log-linear utility. Total land, housing land and transport cost functions are kept general and conditions for the homothetic scaling of the population density profile are derived. Fourth, we analyse whether the homothetic scaling is compatible with a system-of-cities where cities of different populations coexist at equilibrium with the same utility level. Finally, we operationalise the model with functional forms to prepare the empirical validation.

%%%%%%%%%%%%%%%%%%%%%%%%%%%%%%%%%%%%%%%%%%%%%%%%%%%%%%%%%%%%%%%%%%%%%%%%%%%%%%
\subsection*{Alonso-LU and homothetic scaling profiles}

The setting is a featureless plain except for a unique Central Business District (CBD), which concentrates all jobs on a point and is accessed by a radial transport system without congestion. Let $r$ be the Euclidean distance to the CBD and $L(r)$ the exogenous land distribution around the CBD. In reality, $L(r)$ is not necessarily a circle of radius $r$ typically because of water bodies (port cities). In our model, whatever the form of $L(r)$, we depart from Alonso by introducing $H(r)$, the share of $L(r)$  that can be used for housing, hence we have an urban land use augmented model, which we name ``Alonso-LU''. In the Alonso standard model $H(r)=1$ (or any other constant), which obviously contrasts with the blue curve in Fig. \ref{fig_empProf}. In Alonso-LU, we impose $H(r)$ as a portion of $L(r)$ and provide its form exogenously. Densities emerge endogenously but are constrained by the available space $H(r)$ which we know is decreasing with $r$ (Fig. \ref{fig_empProf}). $H(r)$ is only used for housing. Its complement $L(r)[1-H(r)]$ cannot be used for housing (natural and semi-natural areas, transport networks, etc.). 

We now introduce scaling laws. Let us denote by $\rho_{N}(r)$ the population density profile and by $H_{N}(r)$ the profile of the share of urban land used for housing for a city of total population $N$. We assume there exists $\alpha$ and $\gamma$ such that population density profiles $\rho_{N}(r)$ scale homothetically in three dimensions with the power $\alpha$ of population $N$, and that housing land radial profiles $H_{N}(r)$ scale homothetically in the two horizontal dimensions with the power $\gamma$ of population.\endnote{Throughout this paper, scaling properties will implicitly refer to scaling with respect to urban population $N$. Thus, indices \guil{$N$} are used to indicate exogenous variables or functions that are assumed to vary with $N$. Accordingly, indices \guil{$1$} are used to indicate the value of those variables for an abstract unitary city of population $1$.\\[-0.5em]}. These homothetic scaling laws can be formalized as:
\begin{align}
\label{eq_empRhoN}
\rho_{N}(r)=	&\,N^{\alpha}\rho_1\!\left(\frac{r}{N^{\alpha}}\right)\ ,\\
\label{eq_empHN}
H_{N}(r)=		&\,H_1\!\left(\frac{r}{N^{\gamma}}\right)\ ,
\end{align}
\noindent where $\rho_1(r)$ and $H_1(r)$ are the population and land use radial profiles of an abstract unitary city of population $N=1$.

According to \cite{lemoy_scaling_2017}, European urban areas obey equations \ref{eq_empRhoN} and \ref{eq_empHN} (up to some fluctuations which are illustrated later in this work) with the exponents $\alpha\simeq1/3$ and $\gamma\simeq1/2$.

%%%%%%%%%%%%%%%%%%%%%%%%%%%%%%%%%%%%%%%%%%%%%%%%%%%%%%%%%%%%%%%%%%%%%%%%%%%%%%
\subsection*{Residential choice and scaling parameters}

Each household in the model requires land $s$ for housing, work in the CBD and consume a composite commodity $z$ that is produced out of the region and imported at constant price. In that context, residential choice depends only on the distance $r$ to the CBD.
 
Households are rational in the sense of \citet[see also \citealp{myerson_game_1997}]{von_neumann_theory_1944} and their utility function $U$ is
\begin{equation}
\label{eq_U}
U(z,s)=(1-\beta)\ln\!\Big(z(r)\Big)+\beta\ln\!\Big(s(r)\Big)\ ,
\end{equation}
\noindent where $z(r)$ is the amount of composite good (including all consumption goods except housing surface) consumed at distance $r$ from the CBD, $s(r)$ is the housing surface\endnote{In the Alonso model, there is no development of land into housing commodities (land development was introduced into the monocentric theory by \citealp{muth_cities_1969}). Hence the housing market is not distinguished from the land market. Throughout this paper, it is referred as the housing market in order to emphasize Alonso's focus on households' choice. Note also that the term \guil{housing} is used in a broad sense without distinguishing, for example, gardens from built space.\\[-0.5em]} at the same distance $r$ and $\beta\in\ ]0,1[$ is a parameter representing the share of income (net of transport expenses) devoted to housing, or the relative expenditure in housing. Note that $\beta$ is assumed to remain constant across cities of different sizes, which is empirically supported \citep{davis_household_2011}.

Equation \eqref{eq_U} is a log-linear utility function, i.e. the logarithmic transformation of the traditional Cobb-Douglas utility function \citep[from][]{cobb_theory_1928}, and gives the same results in the present case since we work with an ordinal utility. We choose it here for several reasons, which also explain why in urban economic literature it is the form of utility function which is used most often. First, it matches the assumption of a \textit{well-behaved utility function}\endnote{Formally, $U$ must be twice continuously differentiable, strictly quasi-concave with decreasing marginal rates of substitution, positive marginal utilities and all goods must be essentials. See \citet[p.311]{fujita_urban_1989}.\\[-0.5em]} \citep[p.12]{fujita_urban_1989}, which is central in the basic monocentric model and ensures that $U(z,s)$ is defined only for positive values of $z$ and $s$. Second, it contains only a single parameter, $\beta$, which can be discussed empirically. Third, $\beta$ is independent of prices, as found in the empirical literature \citep{davis_household_2011}. Generalization to more general representations of preferences, such as utility functions with constant elasticity of substitution (CES), is left for further studies\endnote{Actually, the log-linear utility is a homothetic function as well since it is the logarithmic transformation of the Cobb-Douglas utility, which is itself homogeneous. This corresponds to a representations of homothetic preferences \citep[see for example][]{varian_introduction_2011}.\\[-0.5em]}.

We choose the composite commodity ($z$) as the numeraire (unit price) and the budget constraint of each household is binding since the households' utility function is monotonic and does not include any incentive to spend money otherwise. The budget constraint at distance $r$ from the CBD is
\begin{equation}
\label{eq_budget}
z+R(r)s(r)=Y_{N}-T_{N}(r)\ ,
\end{equation}
\noindent where $R(r)$ is the housing rent at distance $r$, $Y_{N}$ is the wage of households, and $T_{N}(r)$ is the commuting cost at $r$.
 
We introduce important new scaling assumptions: wages and transport costs are assumed to depend on the total population $N$ of the city. Their variations with city size will strive to reproduce the empirical radial profiles of small and large cities. The measure of agglomeration economies and costs through elasticities of wages and transport costs is well established in the empirical economic literature \citep{rosenthal_evidence_2004, combes_estimating_2010, combes_identification_2011, combes_costs_2012}. This implies power law functions, which are also most often used in urban scaling laws literature \citep{bettencourt_growth_2007, shalizi_scaling_2011, bettencourt_origins_2013, leitao_is_2016}.

Following both strands, we introduce power laws, such that $Y_{N}=N^{\phi}Y_1$, where $Y_1$ is the wage in a unitary city, and $\phi$ is the elasticity of wage with respect to urban population. Similarly, we assume that the transport cost function $T_{N}(r)$ is a scaling transformation (not necessarily homothetic) of $T_1(r)$, the transport cost function in a unitary city (assumed to be continuously increasing and differentiable in $r$). The exact form of this transformation will derive from the conditions for a homothetic scaling, as will be clarified in the next subsection (equation \ref{eq_homT}).

The households' problem consists in maximizing their utility \eqref{eq_U} such that the budget constraint \eqref{eq_budget} holds. 

%%%%%%%%%%%%%%%%%%%%%%%%%%%%%%%%%%%%%%%%%%%%%%%%%%%%%%%%%%%%%%%%%%%%%%%%%%%%%%
\subsection*{Intra-urban equilibrium}

Consider a closed urban region with population $N$. Solving the maximisation problem of households yields the bid rent function $\Psi(r,u)$ (see appendix \ref{apd_households}), which is the maximal rent per unit of housing surface they are willing to pay for enjoying a utility level $u$ (exogeneous) while residing at distance $r$.

Closing the model by linking the utility level $u$ to the population size $N$ yields two additional conditions. First, the quantity of land $L(r)$ at each commuting distance $r$ is finite. Then, summing the population density over the whole (finite) extent of the city, up to the fringe $f_N$, must yield the total population $N$. Second, this fringe is determined by a competition between urban (i.e. housing) and agricultural (default) land uses. We suppose that rents are caught by absentee landowners and that agricultural rent is null for mathematical convenience (see appendix \ref{apd_eq}). This assumption is common in urban economic theory \citep{fujita_urban_1989} and is empirically supported by the low values of agricultural rents relative to housing rents \citep{chicoine_farmland_1981}. Consequently, the urban fringe $f_N$ is the distance at which households spend their entire wage in commuting (and pay a null rent):
\begin{equation}
\label{eq_eqFringeAnull}
f_{N}=T_{N}^{-1}(Y_{N})\ \Leftrightarrow\ Y_{N}=T_{N}(f_{N})\ .
\end{equation}

Finding the unique equilibrium utility and urban fringe satisfying the equilibrium conditions\endnote{To get more information on equilibrium conditions, existence and uniqueness, see \cite{fujita_urban_1989}.\\[-0.5em]} yields the equilibrium population density function $\rho_{N}(r)$ (appendix \ref{apd_eq}). Its homotheticity relies on the three following conditions (appendix \ref{apd_scl}).
\begin{alignat}{3}
\label{eq_homL}
\,\forall\lambda\in\R:\quad		&&\,L(\lambda r)
								&\,=\lambda L(r)\ ,\\
\label{eq_homAlpha}
								&&\gamma
								&\ =\alpha\ ,\\
\label{eq_homT}
\exists\theta\in\R:\quad		&&\,T_{N}(r)
								&\,=N^{\theta}T_1\!\left(\frac{r}{N^{\alpha}}\right)\ .
\end{alignat}

\noindent Condition \eqref{eq_homL} is simply the linearity of $L$, which is clearly satisfied in a two-dimensional circular framework (where $L(r)=2\pi r$). Condition \eqref{eq_homAlpha} states that the horizontal scaling exponents of the share of housing land and population density profiles of equations \eqref{eq_empRhoN} and \eqref{eq_empHN} must be the same. Finally, condition \eqref{eq_homT} actually refines the power-law form of the transport cost function by specifying that the transport cost function is at least (since $\theta$ can be zero) horizontally scaling with power $\alpha$. If these three assumptions hold, then the equilibrium population density function writes
\begin{equation}
\label{eq_eqRhoN}
\rho_{N}(r)=N^{1-2\alpha}H_1(r_{1})\Big[T_1(f_{1})-T_1(r_{1})\Big]^{1/\beta-1}\left[\int\limits_{0}^{f_{1}}L(r_{1})H_1(r_{1})\Big[T_1(f_{1})-T_1(r_{1})\Big]^{1/\beta-1}\dif{r_{1}}\right]^{-1}\ ,
\end{equation}
\noindent where $r_{1}=r/N^{\alpha}$ and $f_{1}=f_{N}/N^{\alpha}$.

We find that this population density profile follows the three-dimensional homothetic scaling \eqref{eq_empRhoN} if and only if $\alpha=1/3$, which coincidentally matches the empirical evidence of \cite{lemoy_scaling_2017}. This means that Alonso's fundamental trade-off between transport and housing is able to explain the observation that cities are similar objects across sizes, provided land profiles, wages and transport costs scale with total population. In other words, a single density profile can be defined from Alonso-LU to match any European city. The main drawback of Alonso-LU is condition \eqref{eq_homAlpha} above, which requires that the scaling exponent of the housing profile is $1/3$, instead of the observed value of $1/2$ \citep{lemoy_scaling_2017}.

%%%%%%%%%%%%%%%%%%%%%%%%%%%%%%%%%%%%%%%%%%%%%%%%%%%%%%%%%%%%%%%%%%%%%%%%%%%%%%
\subsection*{Inter-urban analysis}

Up to now, a closed city of size $N$ has been considered. Yet cities belong to an urban system where households may move from one city to another. This perspective holds two implications. First, since cities of different population size coexist in real urban systems, the equilibrium of the model should be able to reproduce this fact. As a consequence, the benefits and costs of urban agglomeration should vary together when population size changes, so as to compensate each other whatever the size of the city. If one force would dominate the other, the urban system would either collapse to a single giant city or be peppered into countless unitary cities. Second, since by definition households' location decisions are mutually consistent at equilibrium, the equilibrium utility level has to be the same whatever the city population $N$. Otherwise, households would have an incentive to move to larger or smaller cities.

To find out whether the inter-urban equilibrium holds, we substitute the power-law expressions of the wage and transport cost function into the boundary rent and total population conditions. Accounting for the equality of equilibrium utilities across cities of different sizes then yields the following two equalities (see appendix \ref{apd_urbSyst})
\begin{equation}
\label{eq_urbSystEq}
\phi=\theta=\frac{\beta}{3(1-\beta)}\ .
\end{equation}

The left-hand side equality in equation \eqref{eq_urbSystEq} implies that the elasticity of wages with respect to urban population ($\phi$) equals $\theta$, which is the elasticity of the transport cost function once it has been horizontally rescaled. Thus, following the approach of \cite{dixit_optimum_1973}, $\phi$ is representative of urban agglomeration economies whilst $\theta$ results from agglomeration costs\endnote{According to \cite{dixit_optimum_1973}, urban size is mainly determined by the balance between economies of scale in production and diseconomies in transport. Yet in a competitive labour market, labour is paid to its marginal productivity, so that wage-elasticity is representative of labour productivity, which capitalises itself different effects of urban economies of agglomeration. Similarly, the elasticity of the horizontally-rescaled transport cost function catches agglomeration diseconomies.\\[-0.5em]}. Hence the condition for several cities of different population to coexist at equilibrium is met. We note that this equality is supported by recent developments in the very limited empirical literature on agglomeration costs \citep{combes_costs_2012}.

\begin{figure}
\centering
\includegraphics{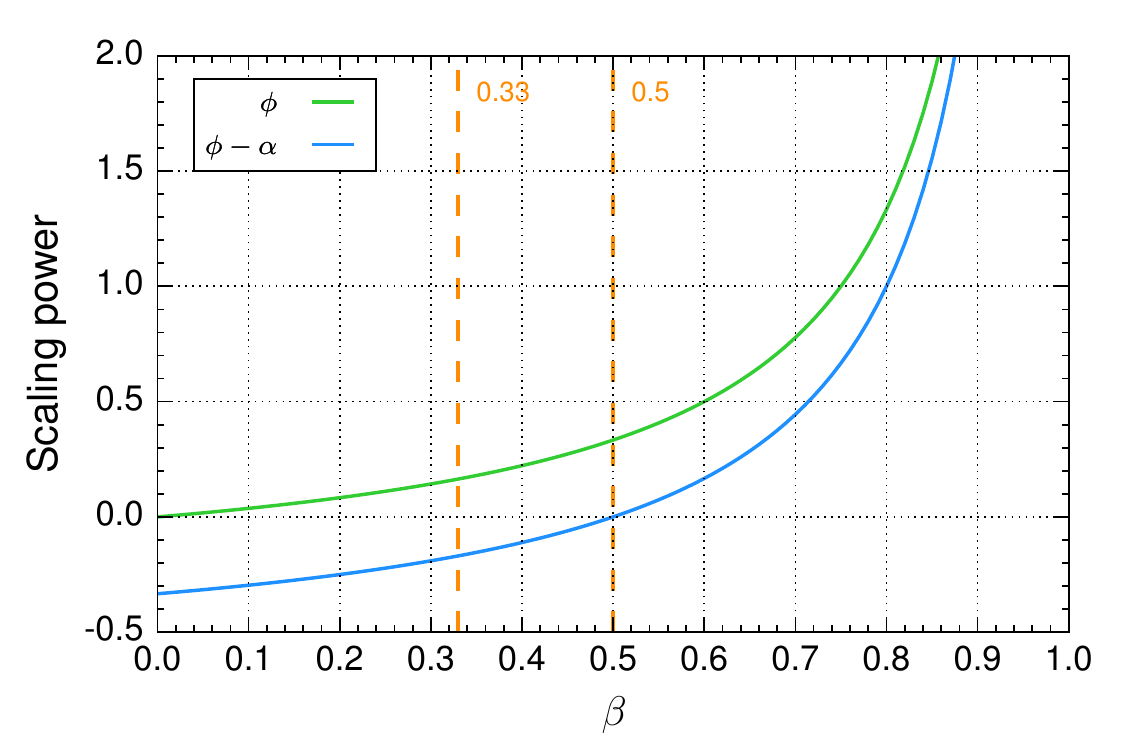}
\captionLine
\caption{\textbf{Population-elasticities of wage and transport cost with respect to housing relative expenditure.} Dashed orange lines highlight values of reference discussed in the text. Recall from equation \ref{eq_urbSystEq} that $\phi=\theta$.}
\label{fig_phiMu}
\end{figure}

The right-hand side equality in equation \eqref{eq_urbSystEq} provides a relationship between the vertical scaling exponent of the value of transport cost $\theta$ (or the population-elasticity of wages $\phi$) and households' relative expenditure in housing $\beta$. This relation is increasing and suggests that a relative expenditure $\beta=1/3$, which is in the range of empirically supported values \citep{accardo_poids_2009, davis_household_2011}, would be associated to exponents $\phi=\theta=1/6$ (Fig. \ref{fig_phiMu}). This value is the same as the superlinearity of socio-economic outputs discussed in \cite{bettencourt_origins_2013, bettencourt_urban_2016}. Consequently, the inter-urban perspective inferred by Alonso-LU is definitely compatible with some former theoretical and empirical researches. However, it diverges from other authors who consider this elasticity to range from $2\%$ to $5\%$ \citep{combes_estimating_2010, combes_identification_2011}. In addition, following other measures of agglomeration economies that are not only based on wages, the elasticity of productivity with respect to city population is considered to be of maximum $3\%$ to $8\%$ \citep{rosenthal_evidence_2004}. Alonso-LU does not solve these empirical incompatibilities. More research effort is needed, especially digging into the functional form of the transport cost function as discussed in the next subsection.

%%%%%%%%%%%%%%%%%%%%%%%%%%%%%%%%%%%%%%%%%%%%%%%%%%%%%%%%%%%%%%%%%%%%%%%%%%%%%%
\subsection*{Functional form}

\paragraph{} We now propose an operational version of the previous model by selecting appropriate functional forms for the land distribution $L(r)$, the housing profile $H_{N}(r)$ and the transport cost function $T_{N}(r)$. The theoretical implications of those forms are discussed as well as their empirical supports. In brief, the functional model is specified by
\begin{align}
\label{eq_funcL}
L(r)=		&\,2\pi r\ ,\\
\label{eq_funcH}
H_{N}(r)=	&\,b\exp\!\left(\frac{-r}{dN^{1/3}}\right)\ ,\\
\label{eq_funcT}
T_{N}(r)=	&\,cN^{\theta - \alpha}r\ , %&\,cN^{(2\beta-1)/[3(1-\beta)]}r\ ,
\end{align}
\noindent where $\theta=\beta/(1-\beta)/3$ (equation \ref{eq_urbSystEq}), $\alpha=1/3$, $b$ is the share of housing land at the CBD, $d$ is the characteristic distance of the housing land profile in a unitary city and $c$ is the transport cost per unit distance in a unitary city. One can easily check that the functional forms \eqref{eq_funcL}-\eqref{eq_funcT} follow the conditions for homotheticity \eqref{eq_homL}-\eqref{eq_homT}, as well as for consistency with the inter-urban approach \eqref{eq_urbSystEq}. The land distribution \eqref{eq_funcL} is simply the usual two-dimensional circular framework and the exponential form \eqref{eq_funcH} of the housing land profile has been chosen for its simplicity and goodness of fit, which is discussed in the next section. 

\paragraph{} We choose a linear form \eqref{eq_funcT} for the transport cost function, because it is largely practiced by urban economists. The elasticity of unitary transport cost with respect to urban population $\theta-\alpha=(2\beta-1)/(1-\beta)/3$ is then endogenously determined by the conditions of homothetic scaling \eqref{eq_homT} and homogeneous utility across cities \eqref{eq_urbSystEq}. It suggests that for $\beta<1/2$ -- which is empirically supported -- the unitary transport cost should decrease with urban population (Fig. \ref{fig_phiMu}). This strives against \cite{dixit_optimum_1973} and the expectation that unitary transport cost is increasing with urban population due to congestion. This shows that the linear transport is not consistent with the scaling of urban profiles. We leave the complete study of a nonlinear transport cost to further research but show in appendix that using a concave transport cost function, which is intuitively more realistic, a positive elasticity of unitary transport cost appears for realistic values of the housing expenditure, e.g. $\beta=1/3$ (appendix \ref{apd_funT}). In particular, changing \eqref{eq_funcT} to $T_{N}(r)=c\sqrt{r}$ (that is, no scaling with population size $N$) would respect our conditions \eqref{eq_homT} and \eqref{eq_urbSystEq}.

\paragraph{} Finally, with the functional form \eqref{eq_funcL}-\eqref{eq_funcT}, the equilibrium population density function \eqref{eq_eqRhoN} becomes (appendix \ref{apd_funRho})
\begin{equation}
\label{eq_funRhoN}
\rho_{N}(r)=\frac{N^{1/3}}{2\pi}e^{-r_{1}/d}\left(f_{1}-r_{1}\right)^{\frac{1}{\beta}-1}{\left[\beta f_{1}^{1/\beta+1}-\left(\beta f_{1}+d\right)e^{-f_{1}/d}d^{1/\beta}\int\limits_{0}^{f_{1}/d}e^{x}x^{1/\beta}\dif{x}\right]^{-1}}\ ,
\end{equation}
\noindent where $r_1=r/N^{1/3}$. This expression depends on three parameters: the unitary urban fringe $f_{1}=Y_1/c$, the housing expenditure $\beta$ and the characteristic distance $d$ of the housing land profile in a unitary city. This density profile model is suitable for empirical calibration. Note that this is a daring exercise since all cities in Europe are calibrated at once using only those three parameters. Its success will expose the descriptive power of the homothetic scaling.

%%%%%%%%%%%%%%%%%%%%%%%%%%%%%%%%%%%%%%%%%%%%%%%%%%%%%%%%%%%%%%%%%%%%%%%%%%%%%%
\section*{Empirics}

\paragraph{} In this section, the functional model \eqref{eq_funRhoN} is calibrated to the average European population density profiles of Fig. \ref{fig_empProf} using nonlinear least squares. The calibration procedure is performed in two steps. First, the optimal value of $d$ is calibrated by comparing the share of housing land \eqref{eq_funcH} to the average profile for a reference city of population $\bar{N}$. Second, the optimal value of $d$ is substituted into the population density function \eqref{eq_funRhoN}, which in turn is calibrated to the average population density profile once by optimizing the values of $f_{1}$ and $\beta$, and once by optimizing only the value of $f_{1}$ with a fixed $\beta=1/3$. Results are visualized for four individual cities.

%%%%%%%%%%%%%%%%%%%%%%%%%%%%%%%%%%%%%%%%%%%%%%%%%%%%%%%%%%%%%%%%%%%%%%%%%%%%%%
\subsection*{Housing land profile}

\paragraph{} We calibrate the share of housing land \eqref{eq_funcH} to the average profile (Fig. \ref{fig_empProf}) for a population of reference $\bar{N}$. This population can be chosen arbitrarily, yet the condition for homothetic scaling imposes a scaling power of $1/3$ which is different from the empirical one ($1/2$). As a result, the model is optimal for the population of reference, but rescaling to other population sizes generates an error. Using the error function detailed in appendix \ref{apd_Nbar}, we choose a population of reference $\bar{N}= 7.03\E{5}$ For a city with this population, the best fit suggests that the characteristic distance is $d=5.8\km$ (Table \ref{tab_bestFit}). Besides, we see that $52.3\%$ of land is dedicated to housing at the CBD, which slightly offsets the average empirical value (Fig. \ref{fig_empProf}). In the Alonso model, the best constant value of housing share is around $17\%$ (Table \ref{tab_bestFit}), which is a poor description of data. 

\begin{figure}
\centering
\includegraphics{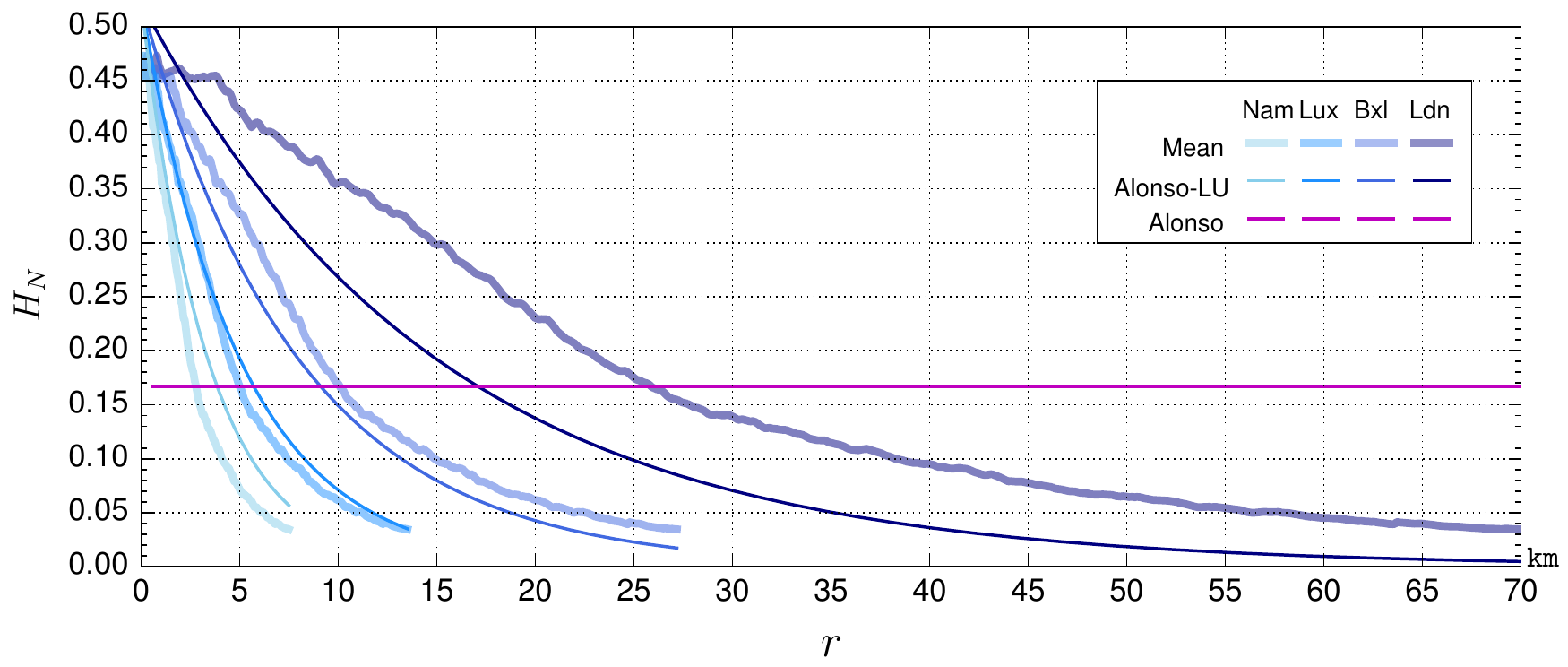}
\captionLine
\caption{\textbf{Calibration of the average housing usage profile.} Mean and best-fit, rescaled to the four example cities.}
\label{fig_housFit}
\end{figure}

\paragraph{} Four cities of different sizes are chosen as illustrations, namely London (\textsf{Ldn}), the largest urban area of the dataset with a population of $N=1.21\E{7}$ in 2006, Brussels (\textsf{Bxl}), the capital of Belgium with $N=1.83\E{6}$, Luxembourg (\textsf{Lux}), capital of the country of the same name, with $N=4.52\E{5}$ and Namur (\textsf{Nam}), the capital of Wallonia in Belgium, with $N=1.39\E{5}$. The population of reference $\bar{N}$, for which the error is minimized, is between those of Luxembourg and Brussels. We see that because of the wrong scaling exponent, the larger the difference between the population $N$ of the considered city and the reference population $\bar{N}$, the larger is the error on housing land share (Fig. \ref{fig_housFit}). For $N>\bar{N}$, the housing share is underestimated, and overestimated for $N<\bar{N}$. In the case of the four considered cities, the absolute error does not exceed $12$ points ($35\%$ in relative terms, see Fig. \ref{fig_housFit}).

%%%%%%%%%%%%%%%%%%%%%%%%%%%%%%%%%%%%%%%%%%%%%%%%%%%%%%%%%%%%%%%%%%%%%%%%%%%%%%
\subsection*{Population density profile}

\begin{figure}
\centering
\includegraphics{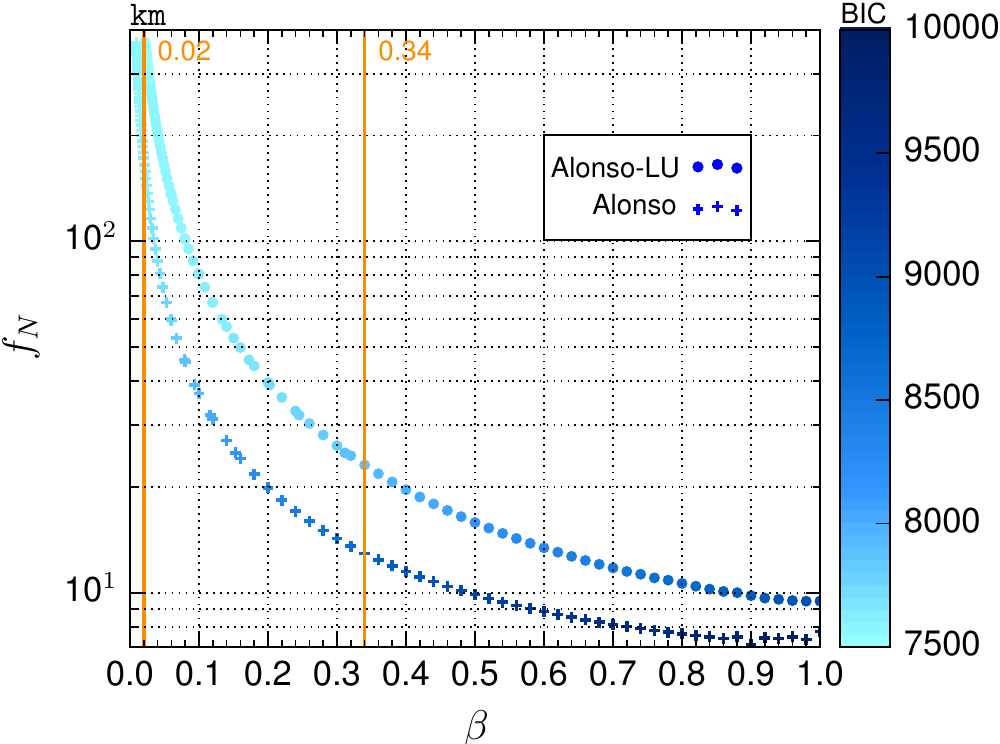}
\captionLine
\caption{\textbf{Best fit parameters for the average population density profile.} The average profile has been rescaled without loss of generality to a reference city of size $\bar{N}=7.03\E{5}$. Colours represent the Bayesian information criterion (BIC). Orange lines show parameter values of the optimal ($\beta=0.02$) and constrained ($\beta=0.34$) models.}
\label{fig_popOptPrm}
\end{figure}

\paragraph{} We turn now to the calibration of the population density function \eqref{eq_funRhoN} with the optimal value $d$ (Table \ref{tab_bestFit}) obtained in the previous subsection to the average population density profile (Fig. \ref{fig_empProf}). We focus again on a city of size $\bar{N}=7.03\E{5}$, this time without loss of generality since the scaling of population density in the model is in agreement with empirical results. The optimal values of the urban fringe $f_{N}$ and of the relative expenditure in housing $\beta$ turn out to be negatively correlated. The best fit is therefore a corner solution with arbitrarily small values of $\beta$ and arbitrarily high values of $f_{N}$ (Fig. \ref{fig_popOptPrm}). In the following we consider the optimal model with $\beta=0.02$. However, this value is unrealistic \citep{davis_household_2011} and thus questions the ability of monocentric models to describe real cities. This issue could probably be solved by including another commuting cost function in our model, but at the expense of mathematical tractability. At this stage, our solution is to also consider a constrained model with $\beta=0.34\simeq1/3$ as a reference case (Fig. \ref{fig_popOptPrm}).

\begin{figure}
\centering
\includegraphics{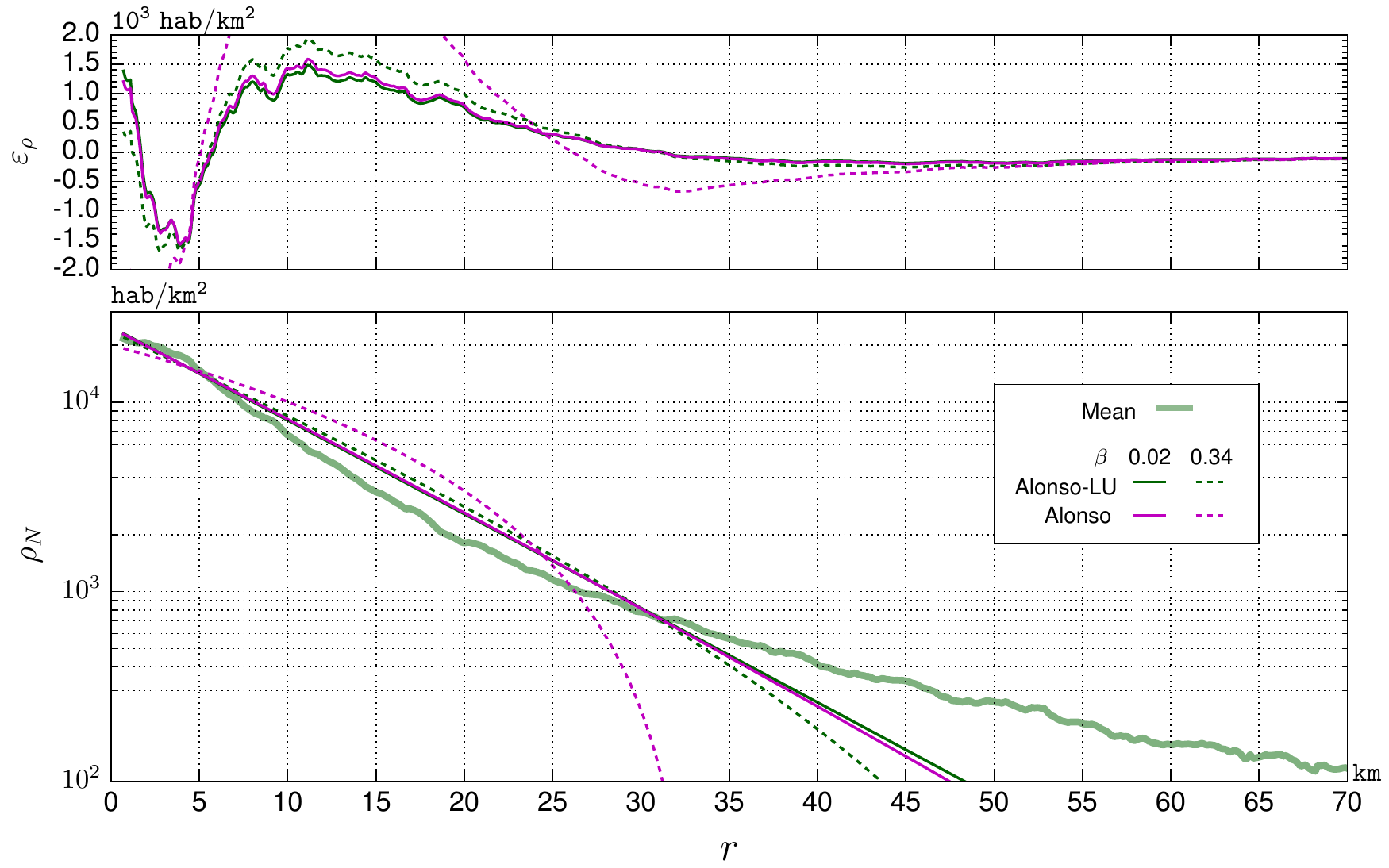}
\vspace{2mm}
\captionLine
\caption{\textbf{Calibration of the average population density profile} Note that the semi-logarithmic scale of the bottom subplot visually exacerbates the error at larger distance from the CBD. On the top subplot, $\varepsilon_{\rho}$ denotes residuals.}
\label{fig_popFit}
\end{figure}

\paragraph{} We look at the best-fit population density profile and focus on the case of London on Fig. \ref{fig_popFit} knowing that smaller cities are obtained by homothetic rescaling. Note that the relative errors are exaggerated because of the semi-logarithmic plot. We observe that the Alonso-LU model outperforms the standard Alonso model, especially for realistic values of $\beta$. Both models display densities whose logarithms are concave because density is going to zero at $r=f_N$. Conversely, the empirical population density profile appears convex. As a result, the best fit model is almost linear in the semi-logarithmic plot (hence almost exponential with linear axes). This form has been long studied empirically in urban economics since \cite{clark_urban_1951}. Theoretical justifications for this exponential form have been provided by \cite{mills_studies_1972,brueckner_note_1982} after adding building construction in the Alonso model, or by \cite{anas_panexponential_2000} who used exponential unitary commuting costs. We contribute a different explanation that is parsimonious and works across city sizes.

\paragraph{} Using the four cities of reference, Fig. \ref{fig_summary} shows that the Alonso-LU model gives a good description of population density profiles for European cities, whatever their size. Four additional cities are provided in appendix \ref{other_cities}, Fig. \ref{fig_summary2}. Visual inspection reveals that the error is mostly due to deviations of individual data from the average profile, and less to deviations of the model from the average profile (Fig. \ref{fig_summary} and \ref{fig_summary2}). We can therefore consider our Alonso-LU model to be very successful.

\paragraph{} Let us note that we do not fix the values of the income $Y_1$ and the unit distance transport cost $c$ in a unitary city since they do not appear in the expression of the population density \eqref{eq_funRhoN}. Our calibration is only performed on land use and population density. We leave to further work a more comprehensive calibration including land prices or rents. Alonso-LU model outputs rent profiles that scale non-homothetically with power $1/3$ in the horizontal dimensions and with power $(1/3+\theta)$ vertically (see Appendix \ref{apd_scl}, also comparison with \citealp{duranton_chapter_2015}). It is flatter than the density profile because the (exponential) $H_N(r)$ factor present in the density disappears in the equation of rents \eqref{eqA_psiN}. This flatter profile seems realistic. However, we do not have radial data for rents across European cities to go further.

%%%%%%%%%%%%%%%%%%%%%%%%%%%%%%%%%%%%%%%%%%%%%%%%%%%%%%%%%%%%%%%%%%%%%%%%%%%%%%
\section*{Conclusion}

\paragraph{} The internal structure of cities obeys a homothetic scaling relationship with total population, which is important to model and explain in order to bridge intra-urban and inter-urban research, and eventually provide new normative hints for urban planning. In this paper, we showed that the fundamental trade-off between transport and housing costs is a good behavioural explanation of this internal structure of cities and holds across city sizes.

\paragraph{} We have proposed an original, augmented version of Alonso's monocentric model (Alonso-LU) that exogenously introduces urban land profile and the scaling of this profile, of wages and transport costs. The model succeeds in reproducing the three-dimensional homothetic scaling of the European population density profiles suggested by \cite{nordbeck1971urban} and recently uncovered by \cite{lemoy_scaling_2017}. Moreover, the model infers the empirical scaling power of $1/3$, and is consistent with an inter-urban perspective, i.e. the coexistence of cities of different sizes. 

\paragraph{} The operationalized version of the Alonso-LU model performs better than the original Alonso model in reproducing the two empirical average profiles. Not only is the fit good, but it is also very parsimonious in parameters (the urban fringe, the housing expenditure, and the decay of the exponential housing land profile). Moreover, comparison with data from individual cities turns out to be surprisingly good in light of the fact that a single parameter (population) is used to adapt the model to different cities.

\paragraph{} Our analysis brings those significant new findings but also comes up with three new challenges. First, the inferred scaling power of the land use profile is significantly smaller than the empirical value of \cite{lemoy_scaling_2017}. Second, we still miss an explanation of this land use profile, which is exogenous here. Third, the proposed model challenges current empirical understanding of wage and transport costs elasticities with population.

\paragraph{} Further research should address those points. In particular, an endogenous model of housing land development is crucial to explain the presence and increase of non-housing land with distance, as well as the scaling of the housing land profile. Potential candidates to this explanation are models of leapfrog urban land development such as \cite{cavailhes_periurban_2004, turner_landscape_2005, caruso_spatial_2007, peeters_emergence_2014} which invoke interaction with agricultural land, or dynamic models with uncertainty like \cite{capozza_stochastic_1990, irwin_interacting_2002}. In the spirit of \cite{muth_cities_1969}, the intensity of housing development (including vertical development) within this urban land should also be addressed in order to better describe cities in their vertical dimension. Finally, the implications of using a nonlinear transport cost need to be addressed in order to shed light on urban agglomeration economies and costs across sizes.

%%%%%%%%%%%%%%%%%%%%%%%%%%%%%%%%%%%%%%%%%%%%%%%%%%%%%%%%%%%%%%%%%%%%%%%%%%%%%%
%\subsubsection*{Acknowledgements}
%
%\paragraph{} The authors are grateful to Jean Cavailhès for insightful comments.

%%%%%%%%%%%%%%%%%%%%%%%%%%%%%%%%%%%%%%%%%%%%%%%%%%%%%%%%%%%%%%%%%%%%%%%%%%%%%%
\theendnotes

%%%%%%%%%%%%%%%%%%%%%%%%%%%%%%%%%%%%%%%%%%%%%%%%%%%%%%%%%%%%%%%%%%%%%%%%%%%%%%
\appendix
\section*{Appendix}

\numberwithin{equation}{section}
\section{Mathematical appendices}

%%%%%%%%%%%%%%%%%%%%%%%%%%%%%%%%%%%%%%%%%%%%%%%%%%%%%%%%%%%%%%%%%%%%%%%%%%%%%%
\subsection{Households consumption problem}
\label{apd_households}

\paragraph{} Taking all the assumptions and notations from the main text as given, households' consumption problem in a city of population $N$ is
\begin{align}
\label{eqA_U}
\max			&\ \Bigg\lbrace U(z,s)=(1-\beta)\ln\!\Big(z(r)\Big)+\beta\ln\!\Big(s(r)\Big)\Bigg\rbrace\\
\label{eqA_budget}
\text{s.t.}		&\ z+R(r)s(r)=Y_{N}-T_{N}(r)\ .
\end{align}

\paragraph{} From the utility function, one computes the marginal rate of substitution
\begin{equation}
\frac{\partial U(z,s)}{\partial z}\left(\frac{\partial U(z,s)}{\partial s}\right)^{-1}=\frac{s(1-\beta)}{z\beta}\ ,
\end{equation}
\noindent which can be equalized to the ratio of prices in order to have the optimal choice equation, that is
\begin{equation}
\label{eqA_optChoice}
\frac{s(1-\beta)}{z\beta}=\frac{1}{R(r)}\ .
\end{equation}

\paragraph{} Simultaneously solving the optimal choice equation \eqref{eqA_optChoice} and the budget constraint \eqref{eqA_budget} by appropriate substitutions yields the solution of the households consumption problem,
\begin{align}
\label{eqA_optZ}
z(r)=	&\,(1-\beta)\Big[Y_{N}-T_{N}(r)\Big]\ ,\\
\label{eqA_optS}
s(r)=	&\,\frac{\beta\Big[Y_{N}-T_{N}(r)\Big]}{R(r)}\ .
\end{align}

\paragraph{} Substituting back the optimal consumptions \eqref{eqA_optZ} and \eqref{eqA_optS} into the utility function \eqref{eqA_U} yields the indirect utility, which can be set to an arbitrary level $u$ in order to express the bid rent function
\begin{equation}
\label{eqA_bidRent}
\Psi(r,u)=e^{-u/\beta}\beta(1-\beta)^{1/\beta-1}\Big[Y_{N}-T_{N}(r)\Big]^{1/\beta}\ ,
\end{equation}
\noindent Finally substituting the bid rent \eqref{eqA_bidRent} into the optimal housing consumption \eqref{eqA_optS} yields the bid-max lot size
\begin{equation}
\label{eqA_bidLot}
s(r,u)=e^{u/\beta}\Bigg[(1-\beta)\Big[Y_{N}-T_{N}(r)\Big]\Bigg]^{1-1/\beta}\ .
\end{equation}

%%%%%%%%%%%%%%%%%%%%%%%%%%%%%%%%%%%%%%%%%%%%%%%%%%%%%%%%%%%%%%%%%%%%%%%%%%%%%%
\subsection{Equilibrium problem}
\label{apd_eq}

\paragraph{} Turning now to the urban equilibrium, let $u$ denote the equilibrium utility level and let $n(r)$ be the population distribution (the population living between $r$ and $r+\dif{r}$) at distance $r$ from the CBD, which is a continuous and continuously differentiable function. Then the following equilibrium relationship states that land available for housing at a given commuting distance $r$ within the city is finite and entirely occupied by households:
\begin{equation}
\label{eqA_noVacant}
L(r)H_{N}(r)=n(r)s(r,u)\ ,
\end{equation}
\noindent where $H_{N}(r)$ follows the horizontal scaling \eqref{eq_empHN}. From this follows the definition of the population density $\rho_N(r)$ in this model:
\begin{equation}
\label{eqA_defdens}
\rho_N(r)=n(r)/L(r)=H_{N}(r)/s(r,u)\ .
\end{equation}

\paragraph{} We express now the two equilibrium conditions. The first one is the boundary rent condition
\begin{equation}
\label{eqA_boundCond}
\Psi(f_{N},u)=a\ ,
\end{equation}
\noindent where $a$ is the exogenous agricultural land rent. As traditionally in urban economic theory, the agricultural land use is no more than a default land use, that is why the agricultural sector is reduced to its most simple form, represented by a constant rent, although it is not really the case empirically \citep{chicoine_farmland_1981, colwell_who_1999, cavailhes_ville_2003}.
The second equilibrium condition is the population condition
\begin{equation}
\label{eqA_popCond}
\int\limits_{0}^{f_{N}}n(r)\dif{r}=N\ .
\end{equation}

\paragraph{} On the one hand, substituting the bid rent function \eqref{eqA_bidRent} into the boundary rent condition \eqref{eqA_boundCond} yields the equilibrium urban fringe
\begin{equation}
\label{eqA_eqFringe}
f_{N}=T_{N}^{-1}\!\left(Y_{N}-(1-\beta)^{\beta-1}\beta^{-\beta}a^{\beta}e^{u}\right)\ .
\end{equation}

\paragraph{} On the other hand, consecutively substituting the optimal housing consumption \eqref{eqA_bidLot} into equation \eqref{eqA_noVacant}, and the resulting value of population distribution into the population condition \eqref{eqA_popCond} yields
\begin{equation}
\label{eqA_eqPop}
e^{-u/\beta}(1-\beta)^{1/\beta-1}\int\limits_{0}^{f_{N}}L(r)H_{N}(r)\Big[Y_{N}-T_{N}(r)\Big]^{1/\beta-1}\dif{r}=N\ .
\end{equation}

\paragraph{} In general, an analytical solution for the equilibrium utility $u$ cannot be obtained by substituting the equilibrium urban fringe \eqref{eqA_eqFringe} into the expression of total population \eqref{eqA_eqPop}. However, with the assumption that the agricultural land rent is null ($a=0$), the equilibrium urban fringe becomes
\begin{equation}
\label{eqA_eqFringeAnull}
f_{N}=T_{N}^{-1}(Y_{N})\ \Leftrightarrow\ Y_{N}=T_{N}(f_{N})\ ,
\end{equation}
 which means that the urban fringe is the distance at which households spend their entire wage in commuting. Equation \eqref{eqA_eqFringeAnull} is very powerful since it enables us to express the results with respect either to the urban fringe $f_{N}$ or to the wage $Y$. It is also linking the two quantities in terms of scaling properties. Now, substituting the right-hand-side equation of \eqref{eqA_eqFringeAnull} into the population constraint yields the equilibrium utility
\begin{equation}
\label{eqA_eqU}
e^{u/\beta}=N^{-1}(1-\beta)^{1/\beta-1}\int\limits_{0}^{f_{N}}L(r)H_{N}(r)\Big[T_{N}(f_{N})-T_{N}(r)\Big]^{1/\beta-1}\dif{r}\ ,
\end{equation}
\noindent which can be consecutively substituted into the optimal housing consumption \eqref{eqA_bidLot} and into equation \eqref{eqA_defdens} in order to express the population density function
\begin{equation}
\label{eqA_rhoN}
\rho_{N}(r)=NH_{N}(r)\Big[T_{N}(f_{N})-T_{N}(r)\Big]^{1/\beta-1}\left[\int\limits_{0}^{f_{N}}L(r)H_{N}(r)\Big[T_{N}(f_{N})-T_{N}(r)\Big]^{1/\beta-1}\dif{r}\right]^{-1}\ .
\end{equation}

\paragraph{}  We note also that the bid rent $\psi_N(r)$ is given by $\psi_N(r)=\beta(T_{N}(f_{N})-T_{N}(r))\rho_N(r)/H_N(r)$, that is
\begin{equation}
\label{eqA_psiN}
\psi_{N}(r)=N\beta\Big[T_{N}(f_{N})-T_{N}(r)\Big]^{1/\beta}\left[\int\limits_{0}^{f_{N}}L(r)H_{N}(r)\Big[T_{N}(f_{N})-T_{N}(r)\Big]^{1/\beta-1}\dif{r}\right]^{-1}\ .
\end{equation}

%%%%%%%%%%%%%%%%%%%%%%%%%%%%%%%%%%%%%%%%%%%%%%%%%%%%%%%%%%%%%%%%%%%%%%%%%%%%%%
\subsection{Conditions of homothetic scaling}
\label{apd_scl}

\paragraph{} In order to derive conditions under which the population density function \eqref{eqA_rhoN} respects the homothetic scaling \eqref{eq_empRhoN}, one first rescales distances accordingly. Formally, under the following change of variable
\begin{equation}
\label{eqA_chngVar1}
r_{1}=\frac{r}{N^{\alpha}}\ ,
\end{equation}
the population density function \eqref{eqA_rhoN} rewrites
\begin{equation}
\label{eqA_rhoN3}
\rho_{N}(r)=\frac{N^{1-\alpha}H_{N}\big(r_{1}N^{\alpha}\big)\Big[T_{N}\big(f_{1}N^{\alpha}\big)-T_{N}\big(r_{1}N^{\alpha}\big)\Big]^{1/\beta-1}}{\int\limits_{0}^{f_{1}}L\big(r_{1}N^{\alpha}\big)H_{N}\big(r_{1}N^{\alpha}\big)\Big[T_{N}\big(f_{1}N^{\alpha}\big)-T_{N}\big(r_{1}N^{\alpha}\big)\Big]^{1/\beta-1}\dif{r_{1}}}\ ,
\end{equation}
\noindent where we note that the urban fringe $f_{N}$ has to be rescaled as well, following
\begin{equation}
\label{eqA_chngVar2}
f_{1}=\frac{f_{N}}{N^{\alpha}}\ .
\end{equation} 
\noindent This has, due to equation \eqref{eqA_eqFringeAnull}, important consequences on the scaling properties of $Y_{N}$ and $T_{N}$.

\paragraph{} Finally assume that $L(r)$ is linearly homogeneous, that $\gamma$ (equation \ref{eq_empHN}), the scaling power of $H_{N}$, is the same as $\alpha$  (equation \ref{eq_empRhoN}), the scaling power of $\rho_{N}$, and that $T_{N}(r)$ is at least horizontally scaling. Formally,
\begin{alignat}{3}
\label{eqA_homL}
\,\forall\lambda\in\R:\quad		&&\,L(\lambda r)
								&\,=\lambda L(r)\ ,\\
\label{eqA_homAlpha}
								&&\gamma
								&\ =\alpha\ ,\\
\label{eqA_homT}
\exists\theta\in\R:\quad		&&\,T_{N}(r)
								&\,=N^{\theta}T_1\left(\frac{r}{N^{\alpha}}\right)\ .
\end{alignat}

\paragraph{} The first assumption will add a \guil{$-\alpha$} term to the power of $N$ in the population density function \eqref{eqA_rhoN3}. The second assumption assumption implies that the horizontal scaling of the housing usage function \eqref{eq_empHN} balances the effect of total population. The third assumption, equivalent to $T_{N}(rN^\alpha)=N^{\theta}T_1(r)$, will enable us to factorize $N^{(1/\beta-1)\theta}$ both in the numerator and the denominator, so that they cancel out. Altogether, this yields
\begin{equation}
\label{eqA_rhoN4}
\rho_{N}(r)=N^{1-2\alpha}H_1(r_{1})\Big[T_1(f_{1})-T_1(r_{1})\Big]^{1/\beta-1}\left[\int\limits_{0}^{f_{1}}L(r_{1})H_1(r_{1})\Big[T_1(f_{1})-T_1(r_{1})\Big]^{1/\beta-1}\dif{r_{1}}\right]^{-1}\ ,
\end{equation}
\noindent which is simply a power function of $N$. In order to finally get the homothetic scaling \eqref{eq_empRhoN} of the population density function, one has to assume that $1-2\alpha=\alpha$ holds, resulting in
\begin{equation}
\label{eqA_eqAlpha}
\alpha=\frac{1}{3}\ .
\end{equation}

\paragraph{} The bid rent can be expressed accordingly as
\begin{equation}
\label{eqA_psiN4}
\psi_{N}(r)=N^{1/3+\theta}\beta\Big[T_1(f_{1})-T_1(r_{1})\Big]^{1/\beta}\left[\int\limits_{0}^{f_{1}}L(r_{1})H_1(r_{1})\Big[T_1(f_{1})-T_1(r_{1})\Big]^{1/\beta-1}\dif{r_{1}}\right]^{-1}\ .
\end{equation}

%%%%%%%%%%%%%%%%%%%%%%%%%%%%%%%%%%%%%%%%%%%%%%%%%%%%%%%%%%%%%%%%%%%%%%%%%%%%%%
\subsection{Consistency with an inter-urban approach}
\label{apd_urbSyst}

\paragraph{} Substituting the scaling of wages into the right-hand-side equation of relationship \eqref{eqA_eqFringeAnull} yields
\begin{equation}
Y_{1}N^{\phi}=T_{N}(f_{N})=T_N(f_1N^\alpha)=N^\theta T_1(f_1)=N^\theta Y_1\ ,
\end{equation}
\noindent where we used also the scalings of the urban fringe \eqref{eqA_chngVar2} and of the transport cost function \eqref{eqA_homT}. This implies
\begin{equation}
\label{eqA_urbSystEq1}
\phi=\theta\ .
\end{equation}

\paragraph{} Second, successively applying the two changes of variable \eqref{eqA_chngVar1} and \eqref{eqA_chngVar2} to the equilibrium utility \eqref{eqA_eqU}, and substituting the conditions of homothetic scaling \eqref{eqA_homL} and \eqref{eqA_homAlpha} yields
\begin{equation}
\label{eqA_eqU2}
e^{u/\beta}=N^{(1/\beta-1)\theta-(1-2\alpha)}(1-\beta)^{1/\beta-1}\int\limits_{0}^{f_{1}}L(r_{1})H_1(r_{1})\Big[T_1(f_{1})-T_1(r_{1})\Big]^{1/\beta-1}\dif{r_{1}}\ .
\end{equation}

\paragraph{} Since at equilibrium households have no incentive to move to another city, equilibrium utility \eqref{eqA_eqU2} should not change with $N$. Thus, equalizing the power of $N$ in equation \eqref{eqA_eqU2} to zero (the rest is independent of $N$) and substituting the value of $\alpha=1/3$ (equation \ref{eqA_eqAlpha}) gives
\begin{equation}
\label{eqA_urbSystEq2}
\theta=\frac{\beta}{3(1-\beta)}\ .
\end{equation}

\paragraph{} Finally, simultaneously solving equations \eqref{eqA_urbSystEq1} and \eqref{eqA_urbSystEq2} yields
\begin{equation}
\phi=\theta=\frac{\beta}{3(1-\beta)}\ .
\end{equation}

%%%%%%%%%%%%%%%%%%%%%%%%%%%%%%%%%%%%%%%%%%%%%%%%%%%%%%%%%%%%%%%%%%%%%%%%%%%%%%
\subsection{Functional transport cost function}
\label{apd_funT}

\paragraph{} Consider the following form of the transport cost function \begin{equation}
T_{N}(r)=cN^{\mu}r^{\sigma}\ ,
\end{equation}
\noindent where $\mu,\sigma\in\R^{+}$. Then the scaling condition \eqref{eqA_homT} requires
\begin{equation}
\label{eqA_funTtheta}
\theta=\alpha\sigma+\mu\ ,
\end{equation}
\noindent where the elasticity $\theta$ of the transport cost function has been broken into two parts. On the one hand, the nonlinear effect of distance contributes by $\alpha\sigma$ to the elasticity $\theta$ because of the horizontal scaling. On the other hand, the contribution of $\mu$ stands for the urban population effects. Further substituting \eqref{eq_urbSystEq} and $\alpha=1/3$ into \eqref{eqA_funTtheta} yields
\begin{equation}
\mu=\frac{\beta-\sigma(1-\beta)}{3(1-\beta)}\ .
\end{equation}

\paragraph{} On the one hand, assuming $\sigma=1$ yields the linear case presented in the functional form \eqref{eq_funcT}. On the other hand, assuming $\beta=1/3$ yields $\mu=1/6-\sigma/3$, which is zero for $\sigma=1/2$.

%%%%%%%%%%%%%%%%%%%%%%%%%%%%%%%%%%%%%%%%%%%%%%%%%%%%%%%%%%%%%%%%%%%%%%%%%%%%%%
\subsection{Functional monocentric model}
\label{apd_funRho}

\paragraph{} Substituting the functional form equations \eqref{eq_funcL}-\eqref{eq_funcT} into the equilibrium population density function \eqref{eq_eqRhoN} with $\alpha=1/3$ yields
\begin{equation}
\label{eqA_funRhoN}
\rho_{N}(r)=\frac{N^{1/3}}{2\pi}e^{-r_{1}/d}\left(f_{1}-r_{1}\right)^{\frac{1}{\beta}-1}\left[\int\limits_{0}^{f_{1}}r_{1}e^{-r_{1}/d}\left(f_{1}-r_{1}\right)^{1/\beta-1}\dif{r_{1}}\right]^{-1}\ ,
\end{equation}

\noindent with $r_1=r/N^{1/3}$ Now, under the change of variable $y=f_{1}-r_{1}$ the integral in equation \eqref{eqA_funRhoN} becomes
\begin{equation}
f_{1}e^{-f_{1}/d}\int\limits_{0}^{f_{1}}e^{y/d}y^{1/\beta-1}\dif{y}-e^{-f_{1}/d}\int\limits_{0}^{f_{1}}e^{y/d}y^{1/\beta}\dif{y}\ ,
\end{equation}
 that the second change of variable $x=y/d$ turns to
\begin{equation}
\label{eqA_intgl}
f_{1}e^{-f_{1}/d}d^{1/\beta}\int\limits_{0}^{f_{1}/d}e^{x}x^{1/\beta-1}\dif{x}-e^{-f_{1}/d}d^{1/\beta+1}\int\limits_{0}^{f_{1}/d}e^{x}x^{1/\beta}\dif{x}\ .
\end{equation}

\paragraph{} The first integral in \eqref{eqA_intgl} can be integrated by parts using
\begin{equation}
x^{1/\beta-1}=\frac{\partial (\beta x^{1/\beta})}{\partial x}\ .
\end{equation}

\paragraph{} After algebraic simplifications, this yields
\begin{equation}
\beta f_{1}^{1/\beta+1}-\left(\beta f_{1}+d\right)e^{-f_{1}/d}d^{1/\beta}\int\limits_{0}^{f_{1}/d}e^{x}x^{1/\beta}\dif{x}\ ,
\end{equation}
\noindent which can be finally substituted to the integral into equation \eqref{eqA_funRhoN} to give
\begin{equation}
\label{eqA_eqRhoN}
\rho_{N}(r)=\frac{N^{1/3}}{2\pi}e^{-r_{1}/d}\left(f_{1}-r_{1}\right)^{\frac{1}{\beta}-1}{\left[\beta f_{1}^{1/\beta+1}-\left(\beta f_{1}+d\right)e^{-f_{1}/d}d^{1/\beta}\int\limits_{0}^{f_{1}/d}e^{x}x^{1/\beta}\dif{x}\right]^{-1}}\ ,
\end{equation}
\noindent and for the bid rent
\begin{equation}
\label{eqA_eqpsiN}
\psi_{N}(r)=\frac{N^{1/3+\theta}}{2\pi}\beta\left(f_{1}-r_{1}\right)^{\frac{1}{\beta}}{\left[\beta f_{1}^{1/\beta+1}-\left(\beta f_{1}+d\right)e^{-f_{1}/d}d^{1/\beta}\int\limits_{0}^{f_{1}/d}e^{x}x^{1/\beta}\dif{x}\right]^{-1}}\ .
\end{equation}

%%%%%%%%%%%%%%%%%%%%%%%%%%%%%%%%%%%%%%%%%%%%%%%%%%%%%%%%%%%%%%%%%%%%%%%%%%%%%%
\subsection{Population of a reference city}
\label{apd_Nbar}

\paragraph{} From equations \eqref{eq_empHN} and \eqref{eq_funcH}, the model of housing usage considered here is a negative exponential with a scaling characteristic distance. Considering the empirical exponents of \cite{lemoy_scaling_2017}, the best model of housing usage is
\begin{equation}
\label{eqA_bestH}
H_{N}(r)=b\exp\!\left(\frac{-r}{gN^{1/2}}\right)\ ,
\end{equation}
\noindent whereas the approximate model is
\begin{equation}
\label{eqA_aprxH}
H_{N}(r)=b\exp\!\left(\frac{-r}{dN^{1/3}}\right)\ .
\end{equation}

\paragraph{} The absolute error between the best model \eqref{eqA_bestH} and the approximate model \eqref{eqA_aprxH} is given by
\begin{equation}
\label{eqA_absErr}
b\exp\!\left(\frac{-r}{dN^{1/3}}\right)-b\exp\!\left(\frac{-r}{gN^{1/2}}\right)=b\exp\!\left(-\frac{r}{gN^{1/2}}\right)\left[\exp\!\left(\frac{-r}{dN^{1/3}}-\frac{-r}{gN^{1/2}}\right)-1\right]\ ,
\end{equation}
\noindent where the relative error is the term between braces. By definition, $\bar{N}$ is a population size chosen arbitrarily, for which the two characteristic distances are equal, thus annihilating the relative error. That is,
\begin{equation}
d=g\bar{N}^{1/6}\ ,
\end{equation}
\noindent such that the relative error rewrites 
\begin{equation}
\label{eqA_hous_errRel}
\exp\!\left(\left[\left(\frac{N}{\bar{N}}\right)^{1/6}-1\right]\frac{-r}{gN^{1/2}}\right)-1\ .
\end{equation}

\paragraph{} It appears from \eqref{eqA_hous_errRel} that for any European city with $N>\bar{N}$, the housing share is underestimated and \textit{vice versa} (Fig. \ref{fig_housFit}). The relative error is bigger, the bigger the difference between $N$ and $\bar{N}$. Hence a first desirable property is that the relative error for the smallest city is the same as for the largest one. This is equivalent to minimizing the maximal relative error. However, this cannot be true for any value of $r$ since the relative error is increasing in $r$. On the opposite, the absolute error \eqref{eqA_absErr} has a maximum value at
\begin{equation}
\bar{r}=-\frac{gN^{1/2}}{6}\ln\left(\frac{\bar{N}}{N}\right)\left[\left(\frac{N}{\bar{N}}\right)^{1/6}-1\right]^{-1}\ ,
\end{equation}

\noindent and at this distance the relative error is simply
\begin{equation}
\left(\frac{\bar{N}}{N}\right)^{\frac{1}{6}}-1\ .
\end{equation}

\paragraph{} Finally, the critical population $\bar{N}$ is chosen as the value for which the absolute value of the relative error at the critical distance $\bar{r}$ is the same for the smallest city in the database, Derry (UK, $1.03\E{5}$\hab), and for the largest, London. This yields
\begin{equation}
\bar{N}=\left(\frac{2}{(1.03\E{5})^{-1/6}+(1.21\E{7})^{-1/6}}\right)^{6}\simeq 7.03\E{5}.
\end{equation}

%%%%%%%%%%%%%%%%%%%%%%%%%%%%%%%%%%%%%%%%%%%%%%%%%%%%%%%%%%%%%%%%%%%%%%%%%%%%%%
\section{Further example cities}
\label{other_cities}
\begin{figure}
\centering
\includegraphics{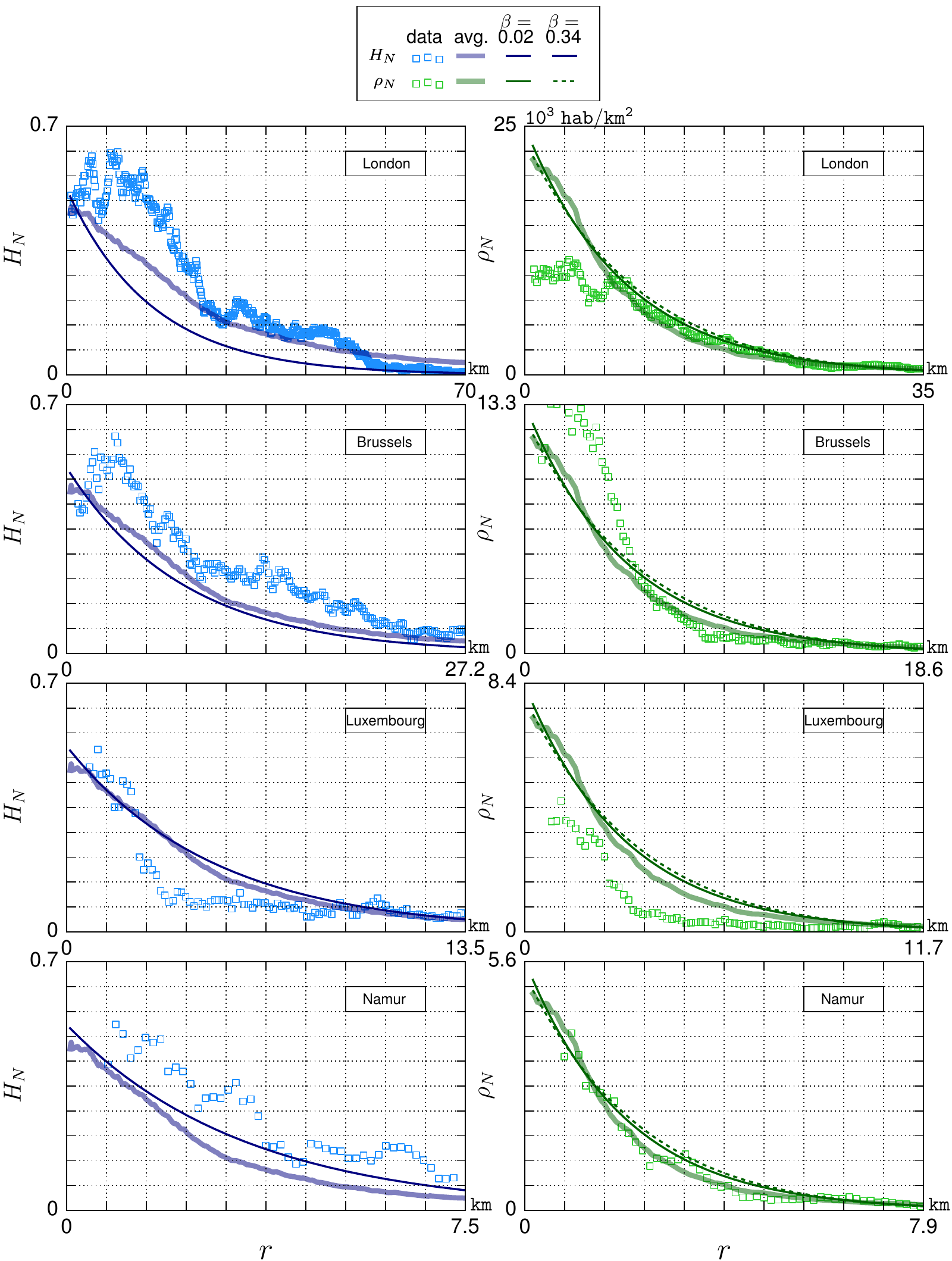}
\vspace{0mm}
\captionLine
\caption{\textbf{Summary plot of the results.} Fitted average profiles compared to individual profiles. Left panel: housing share profile. Right panel: population density profile. Axes have been rescaled to maintain the average curves at the same position across subplots.}
\label{fig_summary}
\end{figure}

\begin{figure}
\centering
\includegraphics{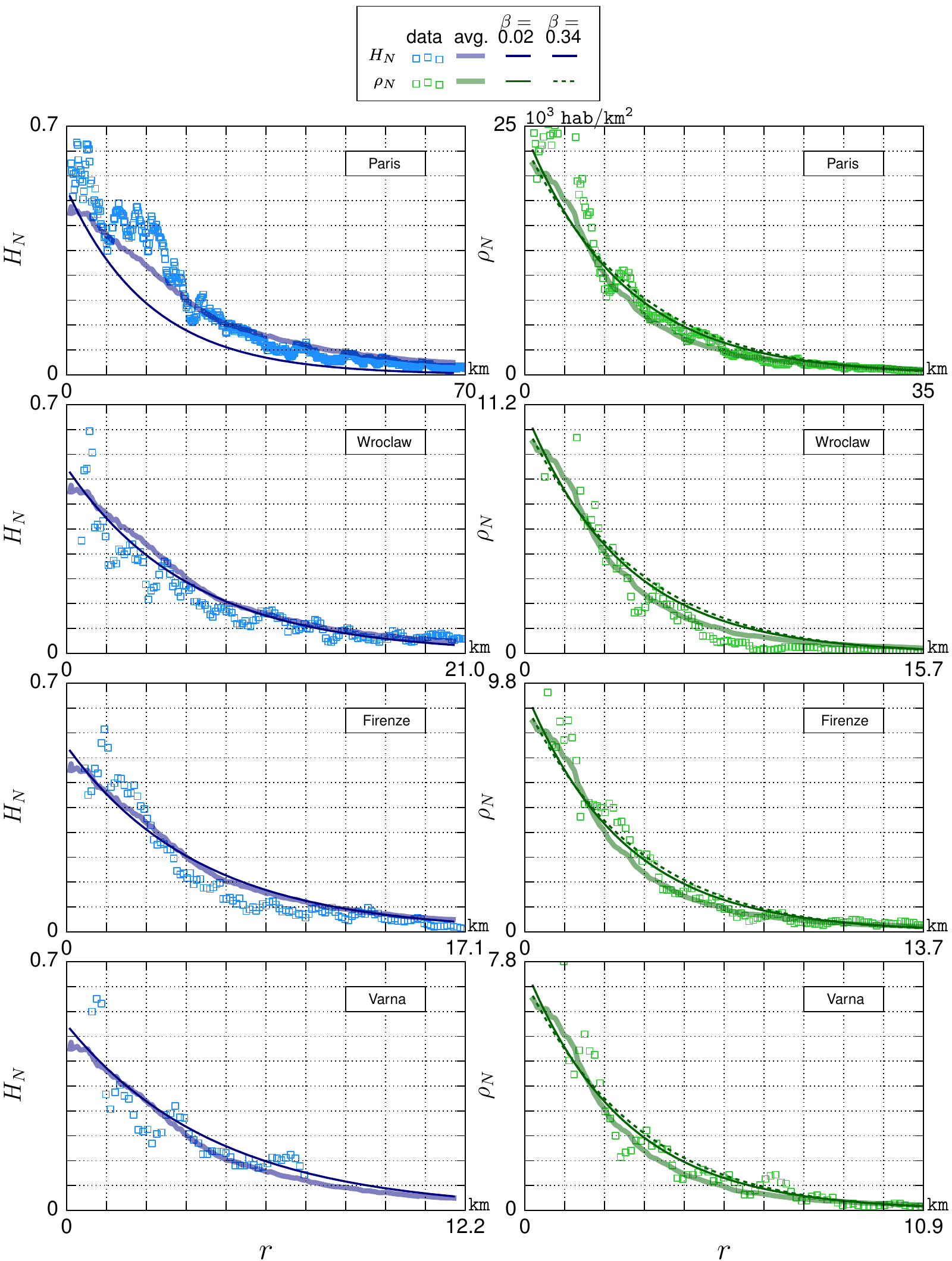}
\vspace{-2mm}
\captionLine
\caption{\textbf{Summary plot of the results.} Fitted average profiles compared to individual profiles. Left panel: housing share profile. Right panel: population density profile. Axes have been rescaled to maintain the average curves at the same position across subplots.}
\label{fig_summary2}
\end{figure}

\paragraph{} We illustrate on Fig. \ref{fig_summary2} the results of the Alonso-LU model on four additional European cities, in order to complement Fig. \ref{fig_summary}: Paris ($N=1.14\E7$), the second biggest city of the database, Wroclaw (Poland, $N=1.03\E6$), Florence (Firenze, in Italy, $N=6.81\E5$) and Varna (Bulgaria, $N=3.48\E5$).

%% references %%%%%%%%%%%%%%%%%%%%%%%%%%%%%%%%%%%%%%%%%%%%%%%%%%%%%%%%%%%%%%%%
\clearpage
\bibliography{bib/myBib3}

%% tables and figures %%%%%%%%%%%%%%%%%%%%%%%%%%%%%%%%%%%%%%%%%%%%%%%%%%%%%%%%
\clearpage
\begin{table*}
\captionsetup{width=.64\textwidth}
\caption{{\bfseries Nonlinear least square results.} Calibration are performed on European average profiles made up with $694$ points, for a population of reference $\bar{N}= 7.03\E{5}$. Distances $d$ and $f_1$ are expressed in kilometers. $C(b,d)$ is correlation between parameters. BIC is the Bayesian information criterion.}
\vspace{-3mm}
\centering
\footnotesize
\sf
\begin{tabular}{>{\raggedleft}p{.075\textwidth}
				>{\centering}p{.05\textwidth}
				>{\centering}p{.08\textwidth}
				>{\raggedleft}p{.05\textwidth}
				>{\centering}p{.05\textwidth}
				>{\centering}p{.05\textwidth}
				>{\centering}p{.05\textwidth}
				p{.05\textwidth}<{\centering}}
\toprule
\multicolumn{3}{l}{A. Housing usage}
		&\multicolumn{5}{l}{B. Population density}\\
		
\cmidrule(l){1-3}\cmidrule(l){4-8}
		&{\raggedright\scriptsize Alonso}
		&{\raggedright\scriptsize Alonso-LU}
		&
		&\multicolumn{2}{l}{\scriptsize Alonso}
		&\multicolumn{2}{l}{\scriptsize Alonso-LU}\\
		
\cmidrule(l){2-2}\cmidrule(l){3-3}\cmidrule(l){5-6}\cmidrule(l){7-8}
$b$				&$0.167$
				&$0.523$
				&$\beta$
				&$0.02$
				&$0.34$
				&$0.02$
				&$0.34$\\[-5pt]
				
				&$(0.005)$	
				&$(0.001)$
				&
				&
				&
				&
				&\\
		
$d$				&$\infty$
				&$5.80$
				&$f_{\bar{N}}$
				&$172.9$
				&$12.94$
				&$409$
				&$23.1$\\[-5pt]

				&
				&$(0.02)$
				&
				&$(0.5)$
				&$(0.07)$
				&$(2)$
				&$(0.2)$\\
		
$C(b,d)$		&
				&$-0.74$
				&
				&
				&
				&
				&\\
		
$\text{BIC}$	&$-2\,808$
				&$-6\,332$
				&$\text{BIC}$
				&$7\,613$
				&$8\,744$
				&$7\,544$
				&$7\,893$\\
\bottomrule
\end{tabular}

\label{tab_bestFit}
\end{table*}

\clearpage

\clearpage

\clearpage

\clearpage

\clearpage

\clearpage

\clearpage

%%%%%%%%%%%%%%%%%%%%%%%%%%%%%%%%%%%%%%%%%%%%%%%%%%%%%%%%%%%%%%%%%%%%%%%%%%%%%%

\end{document}